# SMC-Last Extracted Photometry

T. A. Kuchar[1], G. C. Sloan[2,3], D. R. Mizuno[1], Kathleen E. Kraemer[1], M. L. Boyer[2], Martin A. T. Groenewegen[4], O. C. Jones[5], F. Kemper[6,7,8], Iain McDonald[9], Joana M. Oliveira[10], Marta Sewiło[11,12,13], Sundar Srinivasan[14], Jacco Th. van Loon[10], and Albert Zijlstra[15]

[1] Institute for Scientific Research, Boston College, 140 Commonwealth Avenue, Chestnut Hill, MA 02467, USA  
[2] Space Telescope Science Institute, 3700 San Martin Drive, Baltimore, MD 21218, USA  
[3] Department of Physics and Astronomy, University of North Carolina, Chapel Hill, NC 27599-3255, USA  
[4] Koninklijke Sterrenwacht van België, Ringlaan 3, B–1180 Brussels, Belgium  
[5] UK Astronomy Technology Centre, Royal Observatory, Blackford Hill, Edinburgh, EH9 3HJ, UK  
[6] Institut de Ciències de l'Espai (ICE, CSIC), Can Magrans, s/n, E-08193 Cerdanyola del Vallès, Barcelona, Spain  
[7] ICREA, Pg. Lluís Companys 23, E-08010 Barcelona, Spain  
[8] Institut d'Estudis Espacials de Catalunya (IEEC), E-08034 Barcelona, Spain  
[9] Jodrell Bank Centre for Astrophysics, University of Manchester, Manchester, M13 9PL, UK  
[10] Lennard-Jones Laboratories, Keele University, ST5 5BG, UK  
[11] Exoplanets and Stellar Astrophysics Laboratory, NASA Goddard Space Flight Center, Greenbelt, MD 20771, USA  
[12] Department of Astronomy, University of Maryland, College Park, MD 20742, USA  
[13] Center for Research and Exploration in Space Science and Technology, NASA Goddard Space Flight Center, Greenbelt, MD 20771, USA  
[14] Instituto de Radioastronomía y Astrofísica, UNAM, Antigua Carretera a Pátzcuaro 8701, Ex-Hda. San José de la Huerta, Morelia 58089, Mich., México  
[15] Jodrell Bank Centre for Astrophysics, The University of Manchester, Manchester, M13 9PL, UK  
Received 2023 November 22; revised 2024 February 2; accepted 2024 February 2; published 2024 March 11

## Abstract

We present point-source photometry from the Spitzer Space Telescope's final survey of the Small Magellanic Cloud (SMC). We mapped nearly 30 deg$^2$ in two epochs in 2017, with the second extending to early 2018 at 3.6 and 4.5 $\mu$m using the Infrared Array Camera. This survey duplicates the footprint from the SAGE-SMC program in 2008. Together, these surveys cover a nearly 10 yr temporal baseline in the SMC. We performed aperture photometry on the mosaicked maps produced from the new data. We did not use any prior catalogs as inputs for the extractor in order to be sensitive to any moving objects (e.g., foreground brown dwarfs) and other transient phenomena (e.g., cataclysmic variables or FU Ori–type eruptions). We produced a point-source catalog with high-confidence sources for each epoch as well as a combined-epoch catalog. For each epoch and the combined-epoch data, we also produced a more complete archive with lower-confidence sources. All of these data products will be made available to the community at the Infrared Science Archive.

*Unified Astronomy Thesaurus concepts:* Small Magellanic Cloud (1468); Infrared photometry (792); Celestial objects catalogs (212)

## 1. Introduction

The Small Magellanic Cloud (SMC) is a nearby, metal-poor dwarf galaxy. Its distance (62.44 ± 0.94 kpc; Graczyk et al. 2020) and metallicity ($Z = 0.1$–0.2 $Z_\odot$; e.g., Russell & Dopita 1992; Choudhury et al. 2018) make it an ideal target for studying the evolution of both the interstellar medium and stars with well-characterized distances in a more primitive chemical environment than the Milky Way. Consequently, the SMC has been targeted by many surveys covering a wide wavelength range, from X-rays to radio (e.g., Zaritsky et al. 2002; Cutri & 2MASS Team 2004; Kato et al. 2007; Udalski et al. 2008; Ita et al. 2010; Cioni et al. 2011; Gordon et al. 2011; Haberl et al. 2012; Meixner et al. 2013; Joseph et al. 2019).

The Spitzer Space Telescope (Werner et al. 2004) covered the SMC in whole or in part in multiple epochs. The Spitzer Survey of the Small Magellanic Cloud (S$^3$MC; Bolatto et al. 2007) mapped 2.8 deg$^2$ in the core of the galaxy (outlined in Figure 1) in all seven photometric filters in the Infrared Array Camera (IRAC; 3.6, 4.5, 5.8, and 8.0 $\mu$m; Fazio et al. 2004) and the Multi-band Imaging Photometer for Spitzer (24, 70, and 160 $\mu$m; Rieke et al. 2004). A second Spitzer program, Surveying the Agents of Galaxy Evolution in the Tidally Stripped, Low Metallicity Small Magellanic Cloud (SAGE-SMC; Gordon et al. 2011), followed. It used the same filters but covered the entire galaxy and its environment, including the bar, wing, and tail (these regions are labeled in Figure 1). Spitzer's Last Look at the SMC (or SMC-Last) duplicated the sky coverage of SAGE-SMC. Both surveyed the same 30 deg$^2$, and both observed in two epochs spaced 3–4 months apart. SMC-Last was completed during Cycle 13, which took place during the warm phase of the Spitzer mission. Thus, the only bands available were for the 3.6 and 4.5 $\mu$m IRAC filters.

The S$^3$MC program observed the core of the SMC (the densest regions of the bar) with IRAC in 2005 May. The two IRAC epochs of SAGE-SMC followed in 2008 June and September, and the two SMC-Last epochs were obtained in 2017 (August–September) and from 2017 November to 2018 February. In between SAGE-SMC and SMC-Last, the SAGE-Var program mapped an area of 2.9 deg$^2$ in the core of the SMC in four epochs from 2010 August to 2011 June during the warm Spitzer mission (Riebel et al. 2015). Thus, SMC-Last extends the temporal coverage in the center of the SMC by Spitzer to nine epochs at 3.6 and 4.5 $\mu$m covering a temporal baseline of over 12 yr. The minimum coverage in the SAGE-

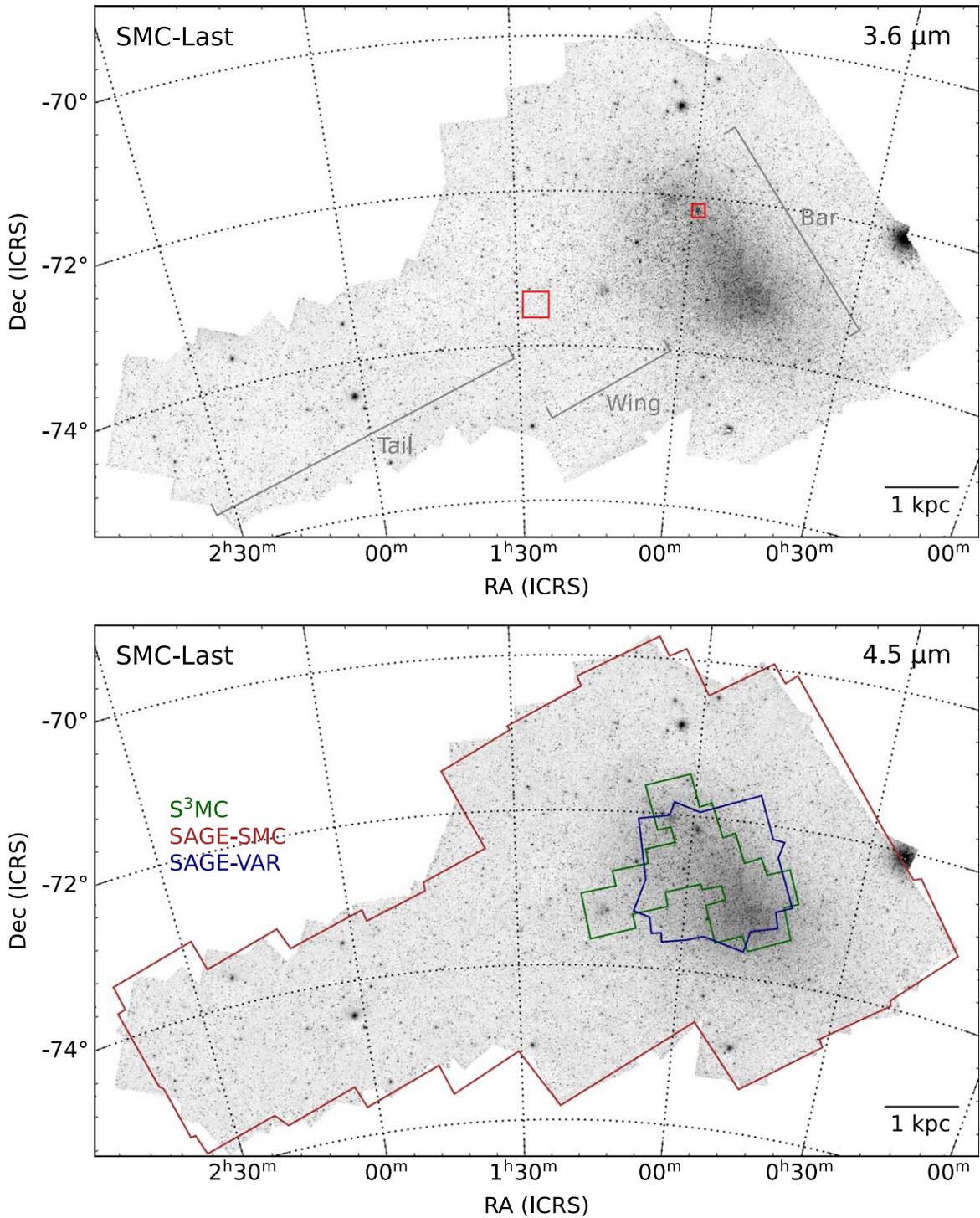

**Figure 1.** The SMC-Last IRAC 3.6 μm (upper panel) and 4.5 μm (lower panel) composite images combining both epochs of observation. The major morphological features of the SMC (the bar, wing, and tail, the latter being a part of the Magellanic Bridge) are labeled. Red squares in the 3.6 μm image indicate the two test fields (the sparse field in the wing and the crowded field centered on NGC 346, as defined in Section 5). In the 4.5 μm image, the colored outlines show the survey areas for S³MC, SAGE-SMC, and SAGE-Var.

SMC footprint is four epochs with a baseline of over 9 yr. Table 1 lists the epochs from Spitzer surveys of the SMC.

The Wide-field Infrared Survey Experiment (WISE; Wright et al. 2010) has provided additional coverage of the SMC in two epochs in 2010 and, starting in 2014, two epochs per year as the Near-Earth Object WISE-Reactivated (NEOWISE-R) mission (Mainzer et al. 2014). WISE scans the entire sky once every 6 months. The first epoch in 2010 was obtained in all four WISE filters. All subsequent epochs followed the loss of cryogens and include just the 3.4 and 4.6 μm filters. The 2023 release of the multiepoch NEOWISE-R catalog includes a total of 20 epochs from 2010 through 2022. While WISE provides better temporal coverage than the Spitzer surveys, it is not as deep, and it is at lower resolution (∼6″, as discussed in Section 5).

Section 2 describes the mosaics from which the point-source catalogs were generated. Section 3 describes the extraction

**Table 1**
Spitzer and WISE Observations of the SMC

| Spitzer Surveys of the SMC | | | | All-Sky WISE[b] |
|---|---|---|---|---|
| Name | Region[a] | Area (deg$^2$) | Epoch | Epoch |
| S$^3$MC | Core of the SMC[c] | 2.8 | 2005 May | 2010 Apr |
| SAGE-SMC | Full SMC | 30 | 2008 Jun | 2010 Oct |
| SAGE-SMC | Full SMC | 30 | 2008 Sep | 2014 May |
| SAGE-Var | Core | 2.9 | 2010 Aug | 2014 Nov |
| SAGE-Var | Core | 2.9 | 2010 Sep | 2015 May |
| SAGE-Var | Core | 2.9 | 2010 Dec | 2015 Nov |
| SAGE-Var | Core | 2.9 | 2011 Jun | 2016 May |
| SMC-Last | Full SMC | 30 | 2017 Aug | 2016 Nov |
| SMC-Last | Full SMC | 30 | 2017 Nov | 2017 Apr |
| … | … | … | … | 2017 Nov |

**Notes.**
[a] The S$^3$MC, SAGE-SMC, and SAGE-Var survey areas are indicated in the 4.5 $\mu$m image in Figure 1.
[b] WISE epochs continue from 2018 to present, two per year.
[c] The "bar" and part of the "wing"; see Figure 1.

process and details how the flux and positional uncertainties were calculated. Section 4 describes the contents of the catalogs and archives. Section 5 compares the SMC-Last point sources to SAGE-SMC and WISE in two test fields. Section 6 summarizes the SMC-Last project.

## 2. Data

The photometry presented here were extracted from mosaic images produced from the SMC-Last survey (Mizuno et al. 2022). The mosaics consist of 52 plates, each 1°.06 × 1°.06, at both 3.6 and 4.5 $\mu$m. These data cover ∼30 deg$^2$ in the SMC, including the main bar of the galaxy, the wing extending to the east, and the tail extending further in that direction toward the Large Magellanic Cloud.

Although the two IRAC bands have slightly different resolutions, the plates were mosaicked to have the same pixel size, 0″.6, and resolution, ∼2″. The observations were taken with high dynamic range exposures at 0.4 and 10.4 s integrations to cover both faint and bright sources and minimize saturation in the images. Two complete maps were made over two epochs separated by 70 days. The first epoch was obtained from 2017 August 25 to September 13 and the second from 2017 November 24 to 2018 February 12. Scheduling logistics required the extension of epoch 2 to 78 days to cover the gaps left in the original scheduling. A third map was generated that combined observations from both epochs, and these data are the basis for Figure 1.

## 3. Source Extraction

We used the open source SExtractor (Source-Extractor version 2.14.2; Bertin & Arnouts 1996) to perform aperture photometry on our mosaicked images. The software is flexible in setting a wide variety of parameters needed for photometry. In general, the software calculates a model of the sky background and then produces a background-subtracted image on which to perform the photometry. Detections are then photometrically extracted, "cleaned" of artifacts (i.e., reexamined for contributions of nearest neighbors), and then saved to a source list. The documentation with the software explains that sources that overlap partially, so that they show two clear peaks, are extracted as separate sources.

### 3.1. Aperture Selection

The selection of aperture size depends largely on the uncertainties expected for the candidate apertures. The primary source of uncertainty for the fainter sources is the background noise in the aperture. For pixel-to-pixel noise $\sigma_{BG}$, the uncertainty in the in-aperture flux will be proportional to $\sqrt{N}\sigma_{BG}$, where $N$ is the number of pixels in the aperture, or equivalently, proportional to $d\sigma_{BG}$, where $d$ is the aperture diameter. The background uncertainty in the total flux, then, is proportional to $C(d)d\sigma_{BG}$, where $C(d)$ is the aperture correction for diameter $d$. We estimate $C(d)$ at small apertures from the 3.6 and 4.5 $\mu$m point-response function (PRF) images supplied by the Infrared Processing and Analysis Center (IPAC).[16] $C(d)$ rises steeply below about 4″ diameter, and we find that the background uncertainty is minimized below a 3″ aperture diameter for both bands.

The second source of aperture-dependent uncertainty is what we term "small-aperture" uncertainty, which arises from the uncertainties in the centroid determination from the extractor, putting varying portions of a source's PRF inside or outside the aperture. This effect increases with decreasing aperture diameter, and it also increases (in absolute flux uncertainties) with flux, because it is a fractional effect, but it is partially mitigated by the result that the centroid uncertainty generally decreases with increasing flux.

Our approach is to apply the aperture that minimizes the total uncertainty for each source individually, particularly with respect to these two competing uncertainties (other sources of uncertainty are either independent of or depend only weakly on the aperture). As the optimum aperture will vary with background noise and source flux, we set the source extractor to report the in-aperture fluxes with a range of aperture diameters: 4″, 5″, 6″, 8″, and 10″. For each aperture, we calculate the total uncertainty and select the results for the aperture that produces the minimum uncertainty. See Section 3.2 for details on the calculation of uncertainties. One exception to this scheme is that for sources with FWHM > 2″.5, or elongation >1.25, a minimum aperture-size criterion is applied regardless of the aperture with the minimum uncertainty, because the aperture corrections (and the resulting uncertainties) are inaccurate for such sources.

The catalog fluxes are the extractor-reported in-aperture fluxes (for the optimum aperture) scaled by the aperture corrections $C(d)$. The aperture correction for the 10″ aperture for each band is interpolated from the correction values shown in the IRAC User's Manual (their Table 4.8 provides cryogenic and warm IRAC aperture corrections). The corrections for the remaining apertures are determined as the median ratio of the 10″ in-aperture fluxes to the in-aperture fluxes at the other apertures over all the epoch 1 sources between 1 and 100 mJy (∼47,000), then scaled to the 10″ aperture correction. We estimate that the uncertainty in the aperture corrections reported in the IRAC User's Manual, about 1%–2%, should hold for the aperture corrections applied for our data. Table 2 shows the aperture corrections we apply for the two bands.

---
[16] https://irsa.ipac.caltech.edu/data/SPITZER/docs/irac/calibrationfiles/psfprf/

**Table 2**
IRAC Aperture Photometry Corrections

| Band | Aperture Size | | | | | |
|---|---|---|---|---|---|---|
| | 10″ | 8″ | 6″ | 5″ | 4″ | 3″ |
| 3.6 μm | 1.070 | 1.100 | 1.148 | 1.205 | 1.329 | 1.649 |
| 4.5 μm | 1.078 | 1.100 | 1.141 | 1.211 | 1.357 | 1.685 |

### 3.2. Uncertainty in Flux

With all such software packages, it is necessary to understand exactly what the software is measuring when it outputs a measurement. For example, the extractor estimates the flux uncertainty given the aperture size(s). However, the uncertainty calculation, particularly for the photoelectron counting statistics, assumes a fixed integration time for the entire image. But since the images are mosaics, they have varying coverages. Also, we have substituted 0.4 s data for the very bright sources that saturated the detectors in the 10.4 s integration. Furthermore, the Spitzer processing pipeline of the raw data includes additional uncertainties, particularly for bright sources, such as uncertainties associated with the nonlinearity correction and correctable saturations. For these reasons, we have chosen to calculate uncertainties separately from the extractor output.

For the aperture photometry, we identify four sources of uncertainty from the pixel values themselves: (1) uncertainty in the background subtracted from the data in the aperture, (2) uncertainty in the aperture integration from pixel-to-pixel rms in the aperture, (3) photoelectron counting uncertainty and other uncertainties for bright sources, and (4) for small apertures, uncertainty in the effective aperture correction due to uncertainties in the positioning of the centroid of the source. For faint sources (below about 0.1 mJy), there is also an uncertainty associated with the correction for a small background-level error.

#### 3.2.1. Background-level Uncertainty ($\sigma_{LEV}$)

The source extractor produces a background rms map. With the parameters of our extractions, this map is generated by determining the rms in $14 \times 14$ pixel bins, with some aggressive outlier deletion. The rms values for each bin are smoothed over 10 bins and then interpolated to give an rms estimate for each pixel. While the rms is formally unique for each pixel, it is basically the average background rms in the surrounding $80'' \times 80''$ region (10 pixels × 14 bins × $0\rlap{.}''6$ pixel$^{-1}$).

For each source, the background rms $\sigma_{\rm BG}$ is determined from the rms map at the source location. The extractor reports the background flux subtracted for the selected aperture, which is converted to a background brightness level by dividing by the aperture area. The background level is determined for each aperture in a 24 pixel wide "square annulus" surrounding the aperture, or roughly a $48 \times 48$ box of pixels. For a flat background, $\sigma_{\rm BG}$ is simply the pixel-to-pixel noise, and the background-level uncertainty would be approximately $\sigma_{\rm BG}/48$. But if $\sigma_{\rm BG}$ contains significant background variations on a length scale larger than the source size but smaller than 14 pixels, then the background uncertainty would be closer to $\sigma_{\rm BG}$ itself. We have no practical way to directly measure background variations around a source. However, as an approximation, we use the background brightness values for each aperture: the rms of the levels over the different apertures is taken, and the greater of this rms (denoted $\sigma'_{\rm BG}$) or $\sigma_{\rm BG}/48$ is used as the background-level uncertainty.

For each aperture, then, the estimated background-level uncertainty is scaled by the number $N$ of pixels in the aperture to give the uncertainty in the in-aperture flux (in integrated MJy sr$^{-1}$ pixel units), then scaled by the solid angle $A$ of the mosaic pixels to give the in-aperture flux uncertainty in mJy ($A = 8.461595 \times 10^{-12}$ sr times the $10^9$ MJy-to-mJy conversion). This result is scaled by the aperture correction $C_k$ for aperture $k$ (Table 2) to give the uncertainty in the total flux.

This uncertainty is adjusted by two additional factors. First, the values in the background rms images are smaller than the expected values of the background rms in the individual IRAC frames divided by the square root of the coverage. Thus, the mosaicking process produces some amount of effective smoothing. This smoothing reduces the measured pixel-to-pixel rms but should not affect the background uncertainties over extended regions, so we scaled the background rms values upward by 20% to approximately match the expected unsmoothed rms in the mosaics. Second, the IRAC frames have a pixel size of $1\rlap{.}''2$, while the mosaics have $0\rlap{.}''6$ pixels, so generating the mosaics effectively rebins the data by a factor of $2 \times 2$. For the uncertainty calculations over ensembles of pixels, this rebinning incorrectly increases the $n^{-\frac{1}{2}}$ rms benefit for sums and averages by a factor of 2, so we scale the calculated uncertainties by 2 to account for the rebinning.

The background-level uncertainty (in mJy) is then

$$\sigma_{\rm LEV} = N(\sigma_{\rm BG}/48)(1.2)(2.0)(A)(C_k).$$

For the case where the empirical background rms $\sigma'_{\rm BG}$ is used, the uncertainty is

$$\sigma_{\rm LEV} = N(\sigma'_{\rm BG})(A)(C_k)$$

because $\sigma'_{\rm BG}$ is a direct measure (albeit an approximation) of the background-level uncertainty without the smoothing and rebinning adjustments.

#### 3.2.2. Aperture Flux Uncertainty Due to Background Noise ($\sigma_{SUM}$)

The background uncertainty $\sigma_{\rm SUM}$ in the aperture integration is also determined from the extractor background rms image. We take the value of the background rms $\sigma_{\rm BG}$ at the source location as constant over the aperture. With this assumption, the uncertainty of the sum of image pixel values over the aperture is $N^{\frac{1}{2}}\sigma_{\rm BG}$, where $N$ is the number of mosaic pixels in the aperture. With the conversion to mJy and applying the adjustments for smoothing and rebinning, we have

$$\sigma_{\rm SUM} = N^{\frac{1}{2}}\sigma_{\rm BG}(1.2)(2.0)(A)(C_k).$$

#### 3.2.3. Photoelectron Counting Uncertainty ($\sigma_{PH}$)

The photoelectron counting uncertainty (and related bright-source uncertainties) is determined from the uncertainty images generated by Spitzer's Mopex mosaicking software (Makovoz & Khan 2005). These images contain the total uncertainty of each mosaic pixel (in MJy sr$^{-1}$) as estimated from the initial processing of the raw IRAC frames. While these uncertainty images contain background uncertainty values, the values do not accurately represent the pixel-to-pixel rms values in our images (Mizuno et al. 2022) and also include overall level uncertainty estimates from the dark subtraction, which do not affect the uncertainty of the aperture photometry. For each source, we subtract the "background" of these uncertainty images by taking a 12 × 12 box of pixels surrounding the source and subtracting the median of the perimeter of the box as the background. Any resulting negative-value pixels are replaced with zeros. The photoelectron counting uncertainty $\sigma_{PH}$ is then

$$\sigma_{PH} = \left(\sum_i p_i^2\right)^{\frac{1}{2}} (2.0)(A)(C_k),$$

where the summation is over the mosaic pixels in aperture $k$ and $p_i$ are the background-subtracted pixel values for the uncertainty image. Note that the mosaic smoothing described above does affect the uncertainty image results but not in a straightforward way, and it is ignored in the present calculation.

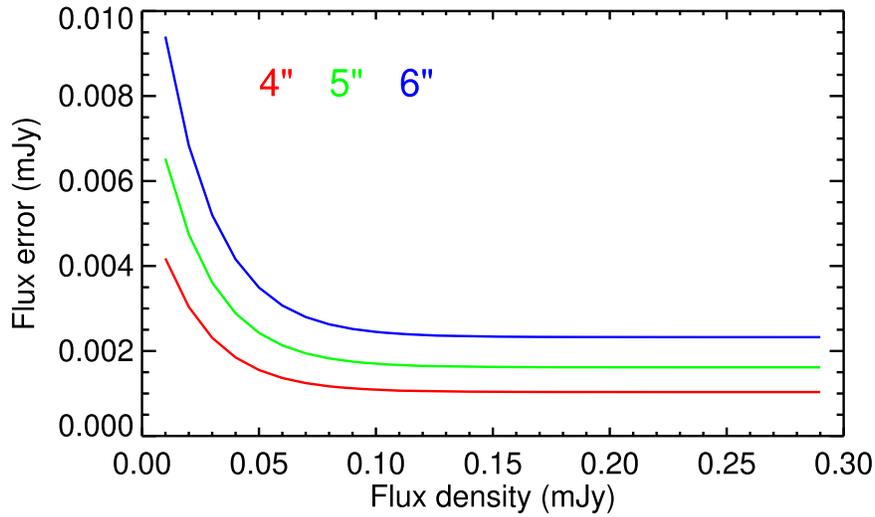

**Figure 2.** Modeled flux corrections due to background-level error for the three smallest apertures and for the 3.6 μm band. The results for the 4.5 μm band are similar.

### 3.2.4. Small-aperture Uncertainty ($\sigma_{SA}$)

For small apertures, uncertainty in the location of the centroid of the source produces additional uncertainty in the calculated flux due to variation in the fraction of the flux contained within the aperture depending on the accuracy of the centroid. We have modeled this small-aperture effect with the PRFs supplied by the Spitzer Science Center. For each band, the average of the 25 images (sampling the focal plane) is taken, with the background subtraction and scaling set to match the aperture corrections supplied in the IRAC User's Manual (Table 4.8). A Monte Carlo simulation of the centroid uncertainties was then performed for each aperture, with the center of the aperture shifted by random amounts in $x$ and $y$ (relative to the PRF center) with a specified rms to simulate the centroid accuracy, and the flux in the aperture was then measured. The resulting scatter in in-aperture fluxes is a model of the fractional uncertainty of the source flux as a function of the aperture size and the centroid rms uncertainty.

For each source, the centroid uncertainty is estimated from an empirical model (see Section 3.3 below) and then applied to the Monte Carlo model to estimate the fractional flux uncertainty for each aperture for that source. These fractional uncertainties are then scaled by the aperture-corrected fluxes (in mJy) to give the uncertainty $\sigma_{SA}$ for each aperture. This uncertainty is a maximum for the faintest sources in the catalog, around 5%–10%, and is a percent or less above about 0.1 mJy.

### 3.2.5. Instrumental Uncertainty ($\sigma_{INST}$)

The IRAC User's Manual (IRAC Instrument & Instrument Support Teams 2021) describes two sources of instrumental uncertainty. The first is the "Array Location" uncertainty and is due to the flat-field correction being performed with the zodiacal background, and so there will be variations with different source spectra. IPAC supplies a correction image (for each band) for a Vega spectrum. We use these to estimate the resulting expected scatter in flux values for our sources by calculating the rms of the correction values in these images. This is an estimate of the uncertainty for a single coverage. Because our data are mosaicked from several IRAC frames, this uncertainty is adjusted by dividing by the square root of the coverage for that source.

The second source is the "Pixel Phase" uncertainty, which arises from different responses of the image pixels depending on the precise location of the source centroid on the pixel. IPAC supplies IDL software to calculate a correction for the location of the source within a pixel.[17] This correction is calculated for a grid of location values on the pixel, and the expected uncertainty is estimated from the rms of the correction values and again scaled for each source by the square root of the coverage.

These two effects differ for each band but together give an expected uncertainty of about 2.4% for both bands. However, these fixed-percentage uncertainties have been adjusted to match an empirical analysis of the uncertainties. See Section 3.2.7 below.

### 3.2.6. Low-flux Background-level Uncertainty ($\sigma_\Delta$)

Systematic variations in the ratios between the fluxes measured at the various apertures (4″, 5″, 6″, and 8″, compared with 10″) for sources with fluxes below ≈0.3 mJy are consistent with a small background-level error, increasing with decreasing flux. Figure 2 shows the resulting modeled flux corrections at the three smallest apertures (the 4″ aperture was used for most of the sources in this flux range). The level error is positive, producing a flux deficit, and so the flux correction is also positive. Note that the background error seems to be a small constant above about 0.3 mJy, but we are omitting the correction when it drops below about 1%. If we denote the correction as $\Delta_k(f)$, for aperture $k$ and flux $f$, then the associated uncertainty for the flux correction is estimated as

$$\sigma_\Delta = \frac{d\Delta_k(f)}{df}\sigma,$$

where $\sigma$ is the flux uncertainty for the source from all other causes.

---
[17] https://irsa.ipac.caltech.edu/data/SPITZER/docs/irac/calibrationfiles/pixelphase/pixel_phase_correct_gauss.pro

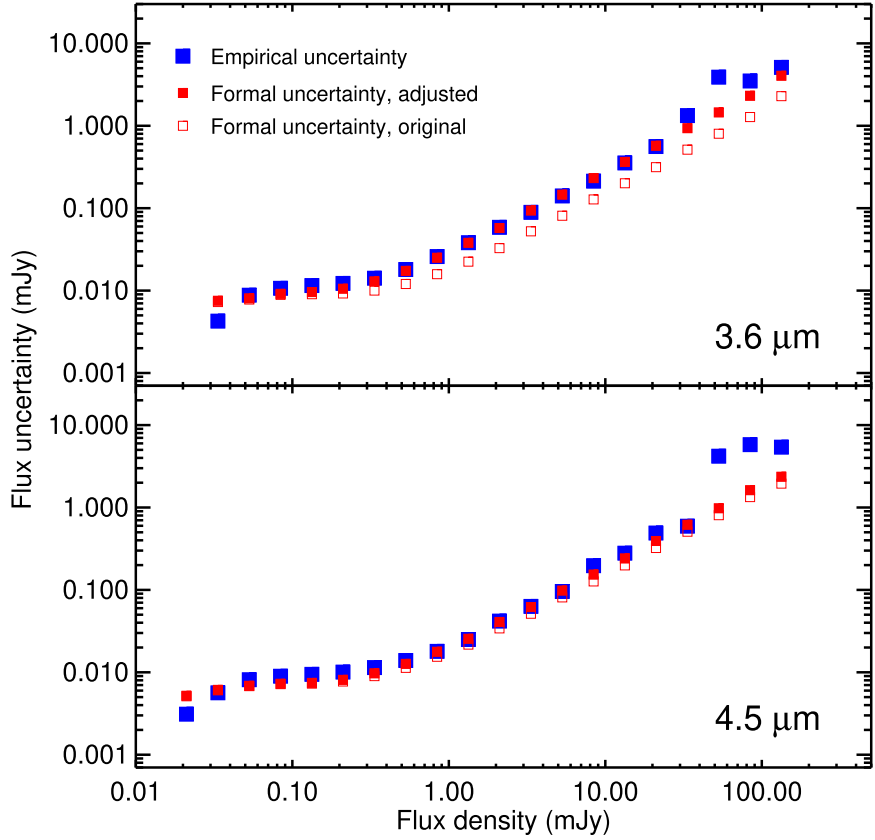

**Figure 3.** Empirical error estimate derived from matching sources across epochs, calculated over flux bins (filled blue squares). For each bin, the rms of the cross-epoch flux differences is calculated and divided by $\sqrt{2}$. The open red squares are the median nominal $\sigma$ formal uncertainty in the bin, and the filled red squares are the $\sigma$ values after the adjustment described in the text.

### 3.2.7. Total Uncertainty and Empirical Analysis

These uncertainties are statistically independent (with the possible exception of $\sigma_\Delta$, as we do not know the mechanism producing the underlying error, and while $\sigma_{\rm LEV}$ and $\sigma_{\rm SUM}$ both depend on the background noise, any associated errors are derived from separate ensembles of pixels). The total uncertainty for a source is then

$$\sigma^2 = \sigma_{\rm LEV}^2 + \sigma_{\rm SUM}^2 + \sigma_{\rm PH}^2 + \sigma_{\rm SA}^2 + \sigma_{\rm INST}^2 + \sigma_\Delta^2.$$

The separate uncertainties are also statistically independent across the two bands. The term $\sigma_{\rm LEV}$ could have some correlation if for a given source that uncertainty is dominated by structured background, which would likely be similar in both bands, but $\sigma_{\rm LEV}$ is never the dominant term in the total uncertainty.

The calculated uncertainties can be checked empirically with the source matches between the two epochs. Noting that the background noise levels at the IRAC frame level for a given band are approximately the same between the two epochs, and considering only sources with two coverages in the mapping scans, the flux uncertainties for a given source between the two epochs should be approximately the same. The flux differences between the epoch-matched sources, then, should in the aggregate reflect the actual uncertainties. Specifically, the rms of the flux differences for an ensemble of sources in a narrow flux range (for which the uncertainties should also generally fall in a narrow range) should be approximately $\sqrt{2}$ times the median uncertainty in the ensemble.

Figure 3 shows the $1/\sqrt{2}$-scaled flux-difference rms values for the epoch-matched sources over a set of flux bins for both bands, here limited to the coverage = 2 sources (filled blue squares). The criterion for inclusion in a flux bin is that the mean of the epoch 1 and epoch 2 fluxes for a given source falls within the flux bin. Also shown are the median formal $\sigma$ (as calculated above) in each bin for the epoch 1 source (open red squares). In the 3.6 $\mu$m band in particular there is clearly a discrepancy between the formal and empirical uncertainty values, the formal being about a factor of 2 too small above a few mJy, but there is also a deficit in the 4.5 $\mu$m band of about 20%–30% (in both cases ignoring the small number statistics at the very high fluxes). Above a few mJy, the only significant contributor to the uncertainty is the instrumental uncertainty (Section 3.2.5), so the discrepancy may be attributed to an inaccurate accounting of that component. Whatever the cause, we have arbitrarily adjusted the instrumental uncertainty to approximately match the empirical results, from 2.4% in both bands to 4.2% and 2.8% in the 3.6 and 4.5 $\mu$m bands, respectively.

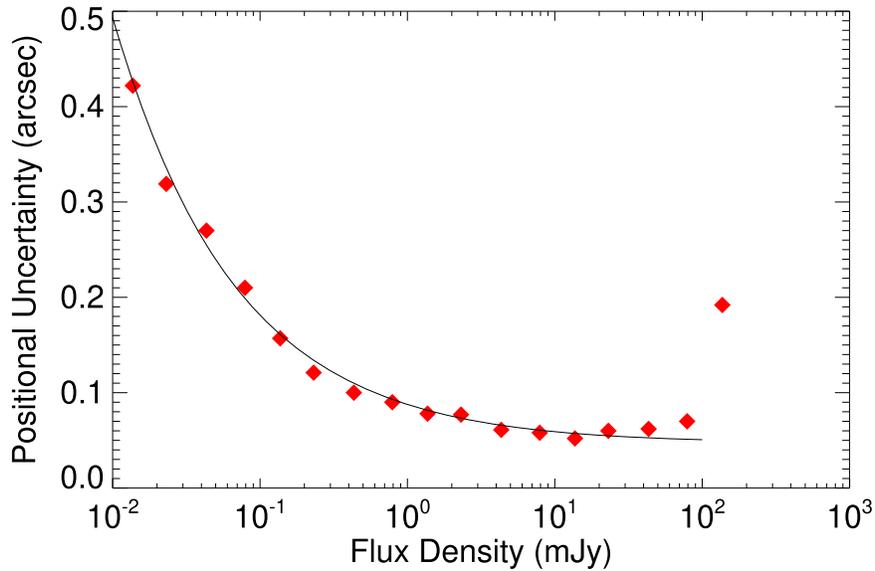

**Figure 4.** Positional uncertainty (single axis) calculated from cross-epoch matches of sources. For each flux bin, the positional differences between epochs in both R.A. and decl. and for both bands are taken. The plotted values are the rms of these bin ensembles divided by $\sqrt{2}$. The epoch 1 3.6 $\mu$m source is used for the flux indexing.

Crowding of sources was not considered in the uncertainty calculations. When another detected source fell within a given source's aperture, that other source was masked (pixels set to 0 after background subtraction), so that to a first approximation, the nearby source does not affect the given source's flux or uncertainty. However, for sources with a very close detected neighbor, less than 3″ or so, we see evidence that the masking does reduce the measured flux, although this effect is not quantified and is not accounted for in the reported uncertainty.

For the most "pointlike" sources (below specified thresholds of FWHM and elongation), the dominant uncertainties are the in-aperture background noise ($\sigma_{\rm SUM}$) for fluxes below about 0.2 mJy. For fluxes above 0.2 mJy, the instrumental uncertainty ($\sigma_{\rm INST}$) dominates. As these are the only two aperture-dependent uncertainties, for the pointlike sources, the flux is the primary determinant of the aperture. For sources that are less pointlike, the aperture is usually increased from the formal uncertainty minimum, largely because the aperture correction is likely to be less accurate for such sources, and a larger aperture should reduce the associated error.

### 3.3. Uncertainty in Position

The extractor determines the position of the source from the pixel position of the emission centroid after background subtraction, which can introduce uncertainties as the extractor subdivides pixels during this process. The accuracy of the measured coordinates of the sources, i.e., the centroid uncertainty in the source extractions, is estimated with the source matches between the two epochs. For each band and the R.A. and decl. separately, the positional differences between the matched sources are taken, and the rms of the differences are calculated over flux bins. If we assume that the positional uncertainties for given sources will be equivalent for both epochs (as both epochs have similar coverage and noise characteristics), then the rms of the differences will be approximately $\sqrt{2}$ times the positional uncertainty of the individual sources (for sources with similar uncertainties). We find that the positional uncertainties are flux-dependent and are essentially equivalent for both bands in both R.A. and decl. Figure 4 shows centroid uncertainties calculated in flux bins.

Figure 4 includes an exponential function fitted to the data of the form

$$\sigma_{\rm CENT}(f) = ae^{-bf} + c,$$

where $f$ is the log of the source flux. This function is the centroid uncertainty used to calculate the small-aperture uncertainty (Section 3.2.4).

In principle, these results include the astrometric uncertainty in the images as well as the centroid uncertainty specifically. However, these data are dominated by sources in regions for which there are hundreds of sources used for the astrometric corrections for each IRAC frame (using matches to the Two Micron All Sky Survey, 2MASS). The IRAC astrometry is corrected (with a translational adjustment) by matching the measured centroids of the sources in the IRAC frame to 2MASS sources (see Appendix B and Mizuno et al. 2022 for details). For that procedure, the centroid accuracy was approximately 0.″5. In the crowded fields, a typical data frame includes about 400 sources and two coverages, so the astrometric accuracy (relative to the mean 2MASS astrometry over the IRAC frame) is about $0.″5/20/1.4 = 0.″018$, generally a small component of the above results. This estimate assumes purely translational uncertainties in the IRAC astrometry relative to the 2MASS astrometry for any given data frame. We do, however, see a small field-rotational error of about 1′ in the 3.6 $\mu$m array astrometry in all the data (see Appendix B). This term gives an effective additional scatter of about 0.″03 in the 3.6 $\mu$m coordinates, which is implicitly included in the $\sigma_{\rm CENT}$ calculation, although it is small enough that differences in positional accuracy between bands can be ignored.

For the positional accuracy reported in the catalog, an adjustment for the astrometric accuracy in the sparser fields is included due to less accurate translational corrections with fewer sources. This is approximately $0.″5/\sqrt{NC}$, where $N$ is the mean number of 2MASS sources used for the astrometric correction and $C$ is the coverage number, so the total positional uncertainty for a source is

$$\sigma_{\text{POS}}^2 = \sigma_{\text{CENT}}^2 + (0\farcs 5)^2/NC.$$

Again, this is relative to the local mean 2MASS astrometry over an IRAC frame.

This positional uncertainty is for a single source detection along a single axis (R.A. or decl.). For sources detected in both bands, the reported R.A. and decl. are the uncertainty-weighted averages of the coordinates determined for each of the bands, which provides roughly a $\sqrt{2}$ benefit in the positional accuracy. The associated single-axis uncertainty of the weighted averages is then scaled upward by $\sqrt{2}$ to give a radial uncertainty.

The reported positional uncertainties are random deviations from the nominal 2MASS astrometry. The absolute uncertainty depends on the accuracy of the 2MASS astrometry in the SMC at the epoch of the SMC-Last observations. The global proper motion of the SMC is approximately 1.5 mas yr$^{-1}$ (Zivick et al. 2018; see also Niederhofer et al. 2021), so for the ~20 yr between the surveys, we expect ~30 mas of positional shift. Also, Zivick et al. (2018) observe proper motions of up to about 0.3 mas yr$^{-1}$ relative to the global SMC proper motion in each of their ~35 separate analysis regions, so this can contribute up to ~6 mas further shift (although Niederhofer et al. 2021 report somewhat higher isolated local proper motions). In addition, we have some evidence of a systematic discrepancy of about $0\farcs 03$ in the measured coordinates of sources between our two epochs, possibly due to a small residual offset error in the pointing corrections, which results in a coordinate shift at different roll angles. The net systematic error we expect is therefore at most $\sim 0\farcs 07$. The brightest objects in the survey, however, are most likely foreground objects and will have much larger proper motions. Caution should be applied with these objects when trying to match across surveys and/or epochs.

## 4. Data Products

### 4.1. Catalogs and Archives

Following the practice of the SAGE project (Meixner et al. 2006), SAGE-SMC (Gordon et al. 2011), and many other Spitzer-based source lists, we have separated the final lists of extractions in a file containing the more reliable sources, referred to as the "catalog," and a more complete "archive," which contains the catalog sources as well as less reliable extractions. Together, these source lists balance completeness and reliability.

Each catalog and archive is published in three versions for a total of six files. We applied the source extractor separately to the mosaics from each observational epoch, and we also extracted sources from the combined-epoch mosaics.

The source extractor generated measured flux densities and preliminary estimated uncertainties for all of the selected apertures. Once the total uncertainties were calculated as described above, the aperture with minimum uncertainty was initially selected for the source list. For sources deviating from "pointlike" (FWHM > $2\farcs 5$ or elongation > 1.25), a larger aperture was generally selected depending on the particular values of FWHM and elongation. Crowding has not been considered for the aperture selection. The source lists include the selected aperture.

For the pointlike sources, below about 2.0 mJy, the in-aperture background noise uncertainty ($\sigma_{\text{SUM}}$) dominates, and the smallest aperture (4″) is usually selected. Above 2.0 mJy, the "small-aperture" uncertainty ($\sigma_{\text{SA}}$) begins to dominate over $\sigma_{\text{SUM}}$ (as it is a fractional uncertainty), and the selected aperture generally increases with increasing flux. Above ~0.3 mJy, the instrumental uncertainty ($\sigma_{\text{INST}}$) dominates the total error, but this uncertainty does not depend on aperture.

The criteria determining whether sources belonged in the catalogs or the archives differ significantly from previous Spitzer-based surveys such as SAGE-SMC. Sources were included in the catalog only if the signal-to-noise ratio (S/N) was 3 or greater in each band. To maintain independence between the observational epochs, sources are not matched between the epochs in either the catalog or the archive. In a given epoch, sources were matched across bands if they were within $1\farcs 6$ of each other. Although most sources in the archive have been matched across bands, that was not a necessary criterion to be in the archive. Therefore, it is possible for an S/N = 3 extraction in IRAC band 1 to be listed in the archive without a corresponding band 2 detection, and vice versa.

We limited the size and shape of the sources to cull extended sources and any remaining artifacts in the images as well as false detections. The extractor uses the distribution of the source pixels to calculate the shape of the source from the second-order moments of the emission. These moments define the ellipticity of the source. We primarily used the FWHM and elongation (ratio of axes). For the catalog, we limited the size and shape of source to remove non-point sources. Sources in the catalogs had to have FWHM ⩽ 3″ and elongations ⩽2. These criteria were relaxed for the archive, with FWHM ⩽ 5″ and elongation ⩽3.5. Thus, source types that can be slightly extended such as young stellar objects are more likely to appear in the archive than the catalog (see also the discussion by Sewiło et al. 2013).

We applied flux criteria to both the catalog and archive for the three epochs. Fainter sources (<0.1 mJy) near brighter ones (>1 mJy) tended to be spurious detections if they were within 12″ of each other (e.g., from diffraction spikes or Airy rings). Therefore, we removed these faint sources from both the catalog and archive. This step culled about 700 sources. Finally, we moved 19 sources that were brighter than 3 Jy at 3.6 $\mu$m to the archive. These objects are badly saturated even in the shorter integration times and thus have less reliable flux densities than required for inclusion in the catalog. The appendix provides a table with these sources.

### 4.2. Modified Julian Date

The catalogs and archives contain a column giving the mean Modified Julian Date (MJD) of the observation, along with maximum and minimum observation times. These values are based on the data for a given epoch and are provided separately for 3.6 and 4.5 $\mu$m. The combined-epoch data also include the mean MJD, calculated in the same manner. The ranges in times are wider than those in a single-epoch catalog, since the starts of the two epochs are separated by 70 days and epoch 2 was observed over a period of 78 days.

### 4.3. Completeness

Figure 5 shows the number of sources binned by magnitude for the two epochs and the combined-epoch catalog for both IRAC bands. To estimate completeness, we extend the linear increase in the log of the number of sources as a function of magnitude and report the magnitude where the cumulative number of sources counted reaches 95% and 90% of the expected total. Table 3 gives the results. The improved S/N in the mosaics for the combined epochs results in significantly better completeness estimates but at the cost of temporal precision.

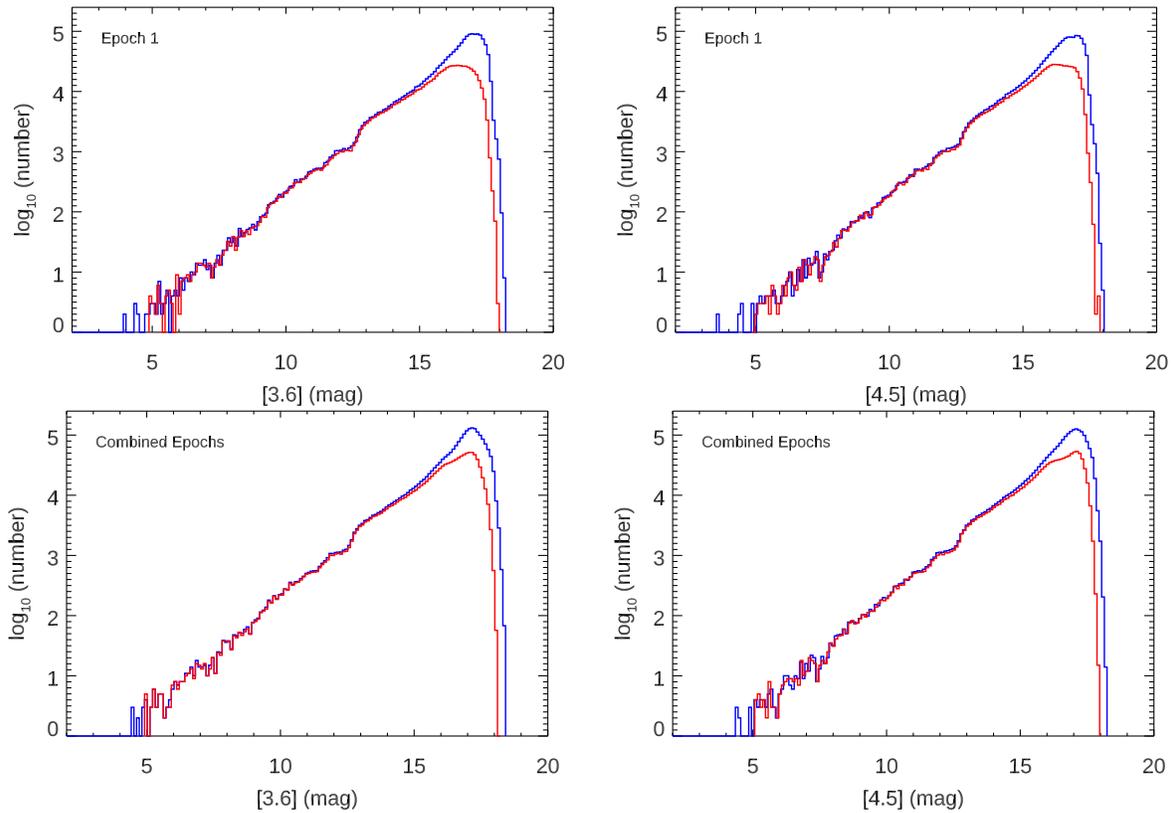

**Figure 5.** Cumulative source counts for the archive (blue) and catalog (red) for epoch 1 and the combined-epoch mosaics at both 3.6 and 4.5 μm. Epoch 1 behaves similarly to epoch 2. Table 3 gives the limiting magnitudes for 95% and 90% completeness.

Table 4 compares SMC-Last to SAGE-SMC, showing the number of sources and minimum and maximum flux densities and magnitudes for the various iterations of the catalogs and archives. Here and in the following sections in this paper, comparisons to SAGE-SMC are based on the SAGE-SMC IRAC "Single Frame + Mosaic Photometry" Catalog and Archive, v1.5,[18] which are based on photometry extracted from the combined epoch 1 and epoch 2 mosaics (Gordon et al. 2011). For both SMC-Last and SAGE-SMC, the archive includes all sources in the catalog. The source counts reveal a difference in how the catalogs and archives were separated in the two surveys. In SAGE-SMC, only about 10% of the sources are unique to the archive, while in SMC-Last, a more conservative approach has led to more sources unique to the archive than are in the catalog. The total source counts for the archives in both surveys are comparable.

## 5. Comparing SMC-Last to Other Mid-infrared Surveys

As outlined in the Introduction, multiple surveys with Spitzer and WISE have covered the SMC. The WISE mission has led to the AllWISE catalog (Cutri et al. 2013), which is based on the original two epochs obtained in 2010, and the CatWISE catalog (CatWISE2020; Marocco et al. 2021), which coadds the first 12 WISE, NEOWISE, and NEOWISE-R epochs (from 2010 through 2018). The AllWISE and CatWISE catalogs require S/N > 5 for inclusion of a source, but only in a single band, as opposed to the SMC-Last catalogs, which require S/N > 3 in *both* bands.

---

[18] Available from the Infrared Science Archive (IRSA); https://irsa.ipac.caltech.edu.

**Table 3**
Limiting Magnitudes and Completeness of the SMC-Last Survey

| Epoch | Product | 3.6 μm | | 4.5 μm | |
|---|---|---|---|---|---|
| | | 95% | 90% | 95% | 90% |
| 1 or 2 | Catalog | 15.6 | 16.2 | 15.8 | 16.4 |
| Combined | Catalog | 15.7 | 16.7 | 16.5 | 16.9 |
| 1 or 2 | Archive | 17.5 | 17.6 | 17.4 | 17.5 |
| Combined | Archive | 17.8 | 17.9 | 17.8 | 17.9 |

The ability to extract a photometric source from the data depends on the complexity of the field. We examined two test fields, as shown in Figure 6 (also indicated in Figure 1). One field, referred to as the "crowded" or "NGC 346 field," is $12' \times 12'$ and centered on NGC 346 ($\alpha, \delta = 14°.7721, -72°.17589$). The other field samples a sparsely populated region $20' \times 20'$ in size and centered at $21°.70, -73°.32$. We refer to it as the "sparse field."

To compare the coverage of these surveys, we performed two tests in the sparse and crowded fields, using the SAGE-SMC IRAC "Single Frame + Mosaic Photometry" Archive, v1.5, as the basis for comparison. First, we extracted all sources from these regions for SAGE-SMC, AllWISE, CatWISE, and all three epochs (1, 2, and combined) of the SMC-Last archive. Table 5 gives the source counts in each field for the SAGE-SMC archive,

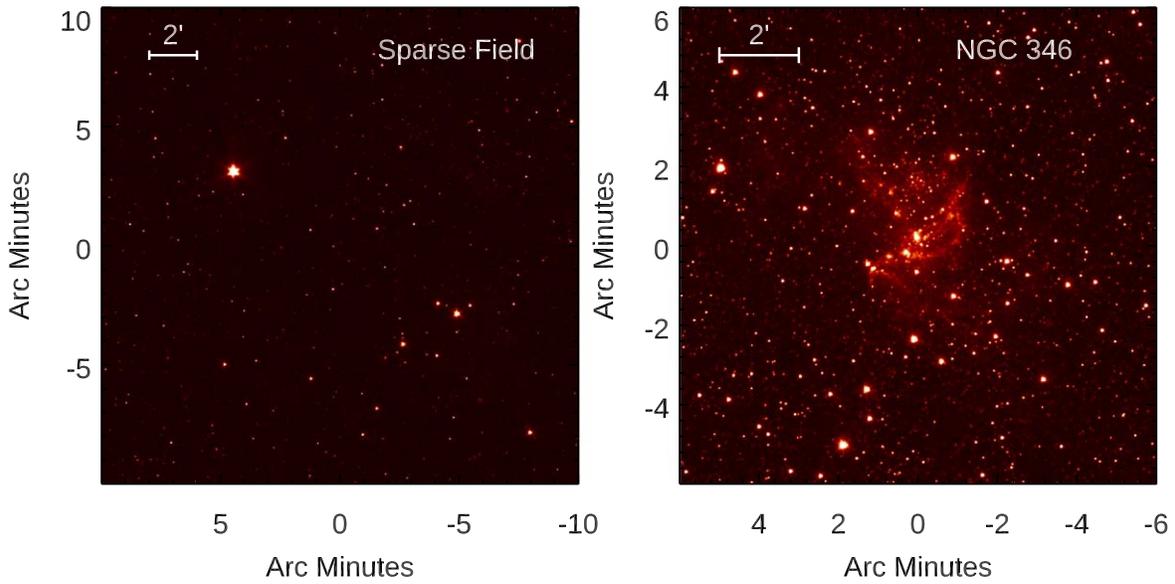

**Figure 6.** Test fields: (left) the $20' \times 20'$ sparse field centered at $(\alpha, \delta) = (21°70, -73°32)$; (right) the $12' \times 12'$ crowded field centered near NGC 346 at $(\alpha, \delta) = (14°77, -72°18)$.

Table 4
Source Statistics for SMC-Last and SAGE-SMC

| Product | Epoch | Number of Sources | Flux Density (and Magnitude) Ranges (mJy and mag) | | | |
|---|---|---|---|---|---|---|
| | | | 3.6 μm | | 4.5 μm | |
| | | | Min. | Max. | Min. | Max. |
| SMC-Last | | | | | | |
| Catalog | 1 | 651,614 | 0.02 (17.76) | 2946 (4.95) | 0.02 (17.48) | 1879 (4.95) |
| | 2 | 683,722 | 0.02 (17.84) | 2925 (4.96) | 0.02 (17.47) | 1844 (4.97) |
| | Comb. | 1,002,360 | 0.02 (17.96) | 2844 (4.99) | 0.02 (17.68) | 1579 (5.14) |
| Archive | 1 | 1,820,897 | 0.01 (18.24) | 71,370 (1.49) | 0.01 (18.03) | 37,360 (1.71) |
| | 2 | 1,850,282 | 0.01 (18.31) | 71,200 (1.49) | 0.01 (18.05) | 36,340 (1.74) |
| | Comb. | 2,397,006 | 0.01 (18.46) | 71,330 (1.49) | 0.01 (18.28) | 37,030 (1.71) |
| SAGE-SMC | | | | | | |
| Catalog | SMP[a] | 2,015,403 | 0.01 (20.12) | 1118 (6.0) | 0.01 (19.60) | 1134 (5.5) |
| Archive | SMP[a] | 2,194,836 | 0.01 (21.05) | 1118 (6.0) | 0.01 (20.37) | 1134 (5.5) |

**Note.**
[a] SMP refers to the SAGE-SMC IRAC "Single Frame + Mosaic Photometry" Catalog and Archive, v1.5. The values cited are the minima and maxima in the databases at IRSA. Note that these differ from those given by Gordon et al. (2011), which are the values where 99% of the sources are brighter than the reported number (SAGE–SMC Team & Gordon 2011) and used in Figure 8.

the two WISE catalogs, and the separate and combined-epoch archives from SMC-Last (in the "No. of Sources" columns).

The source counts show that the crowded nature and complex backgrounds in the NGC 346 field strongly affect the WISE data, which have lower angular resolution than Spitzer. Both Spitzer surveys, SAGE-SMC and SMC-Last, have a resolution of $\sim 2''$, compared to $\sim 6''$ for WISE (Wright et al. 2010). In the sparse field, CatWISE outperforms the Spitzer surveys, showing the power of coadding the available WISE epochs.

The second test was to match sources in the surveys to SAGE-SMC to compare the source recovery levels as a function of magnitude. WISE sources had to be within $2''5$ of the SAGE-SMC source, and SMC-Last sources had to be within $1''5$. All sources had to be within 2 mag of the SAGE-SMC source. Table 5 also shows, for both test fields, the magnitude limit at which 95% and 90% of the SAGE-SMC sources can be recovered in all of the surveys. For SAGE-SMC, the limits mark the magnitudes at which the sample reaches that percentage of its total number of recovered sources.

It should be noted that the magnitude limits in Table 5 are not true completeness limits, because the SAGE-SMC archive can only approximate the actual population of sources in the test fields for a variety of reasons. Each of these surveys estimated magnitude limits to some extent, but they used different methodologies. The estimates in Table 5 are calculated identically, making them more comparable to each other. Also, the survey completeness limits are calculated for our two test fields, rather than for the entire survey. As discussed above, the magnitude limits depend on the complexity of the field, with more complex fields leading to shallower magnitude limits. Nonetheless, the magnitude limits given do allow some useful comparisons.

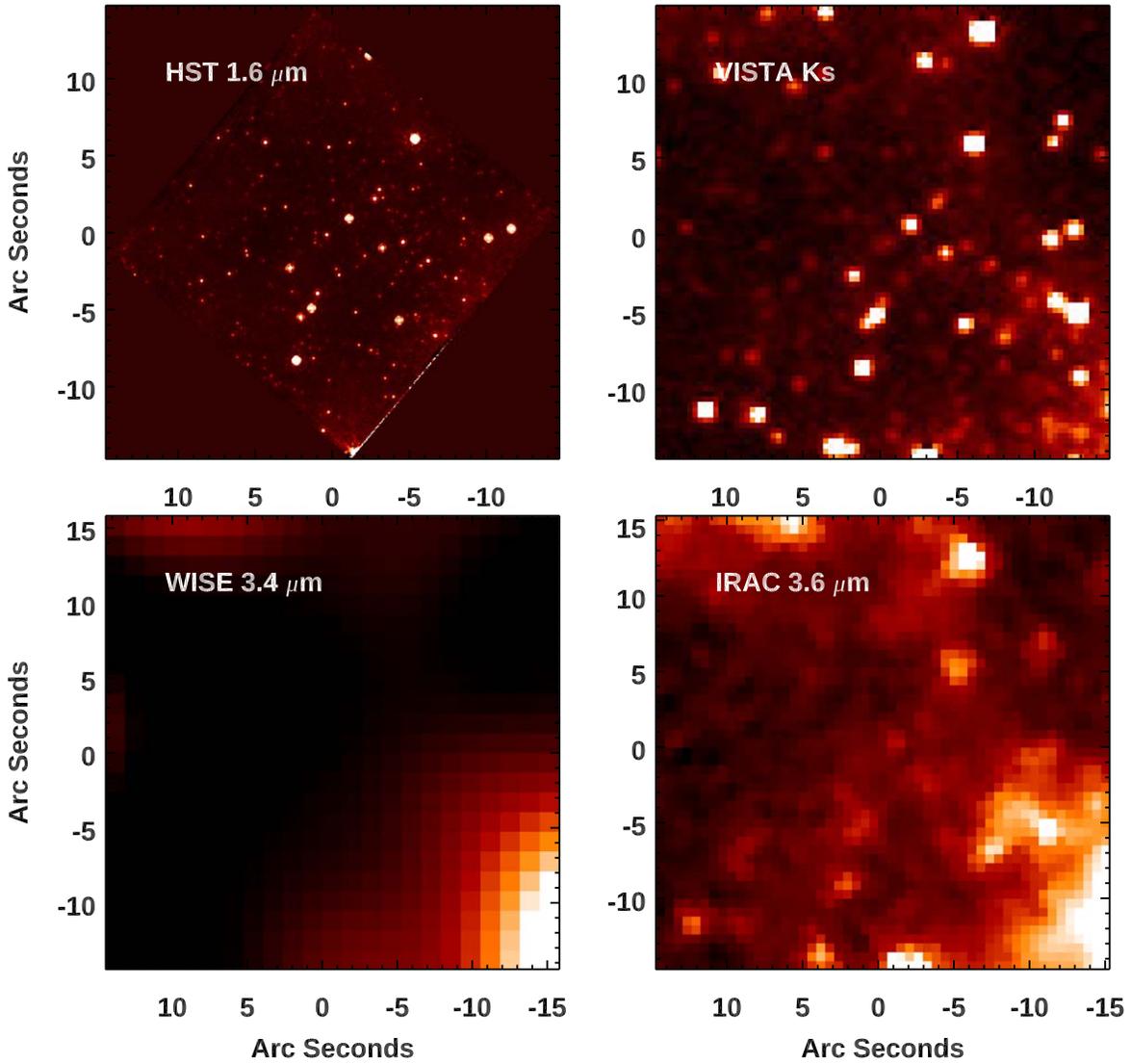

**Figure 7.** A crowded field near NGC 346. (Upper left) 1.6 μm data from NICMOS on Hubble; (upper right) Ks data from the VISTA 4 m telescope; (lower left) 3.4 μm data from AllWISE; (lower right) 3.6 μm data from SMC-Last. This ∼30″ × 30″ region is located slightly northeast of the center of NGC 346 in Figure 6 at $(\alpha, \delta) \approx (14\overset{\circ}{.}79, -72\overset{\circ}{.}17)$ (the only location within our test fields with Hubble near-infrared imagery).

As with the total source counts, the limitations of the WISE-based catalogs in crowded fields are readily apparent in the vicinity of NGC 346. The magnitude limits for AllWISE give a good idea of the depth of the individual epochs in the NEOWISE-R catalogs. In uncrowded fields, these epochs are a useful addition to the epochs provided by SAGE-SMC and SMC-Last, but in crowded fields, many sources will be missing or confused. Figure 7 compares near-infrared data near NGC 346 with higher angular resolution from Hubble (∼0″.1) and VISTA (∼0″.5) to the same field from AllWISE (∼6″) and SMC-Last (∼2″).

The stringent requirements for the SMC-Last catalog lead to limitations similar to AllWISE, with crowding definitely impacting what can be recovered. The archive is significantly deeper in these fields, but users should check the various flags to ensure the data they are using match their needs.

The same caveat applies to the choice of SMC-Last archive versus catalog in all fields, and reference to Table 5 can indicate which source of data users should rely on depending on the expected magnitude of their targets of interest and the crowdedness of the field.

Figure 8 shows the color–magnitude diagrams for the two test fields for both SMC-Last and AllWISE. The figure contains the magnitude/color limits for 95% matches of the AllWISE, CatWISE, and SAGE-SMC surveys for the test fields, as defined above and given in Table 5. These limits should not be confused with sensitivity limits. For example, SAGE-SMC contains sources down to 20th magnitude, as given in Table 4. Table 5 shows that the limits of the CatWISE and SMC-Last surveys are similar in the sparse field, where the multiple epochs compensate for the lower sensitivity of WISE. However, as Figure 8 shows, the crowded nature of the NGC 346 field has impacted both of the WISE-based surveys due to the lower angular resolution of WISE.

Since the pointing of the mosaic images was updated using the 2MASS catalog (for details, see Appendix B and Mizuno et al. 2022), it was worthwhile to compare the consistency of the SMC-Last catalog positions in the two test fields with those of SAGE-SMC, which was also referenced to 2MASS. We limited matches between the two surveys to sources that were within 2″ of each other. We identified ∼1500 matched sources in the NGC 346 field and ∼1200 in the sparse field. More than

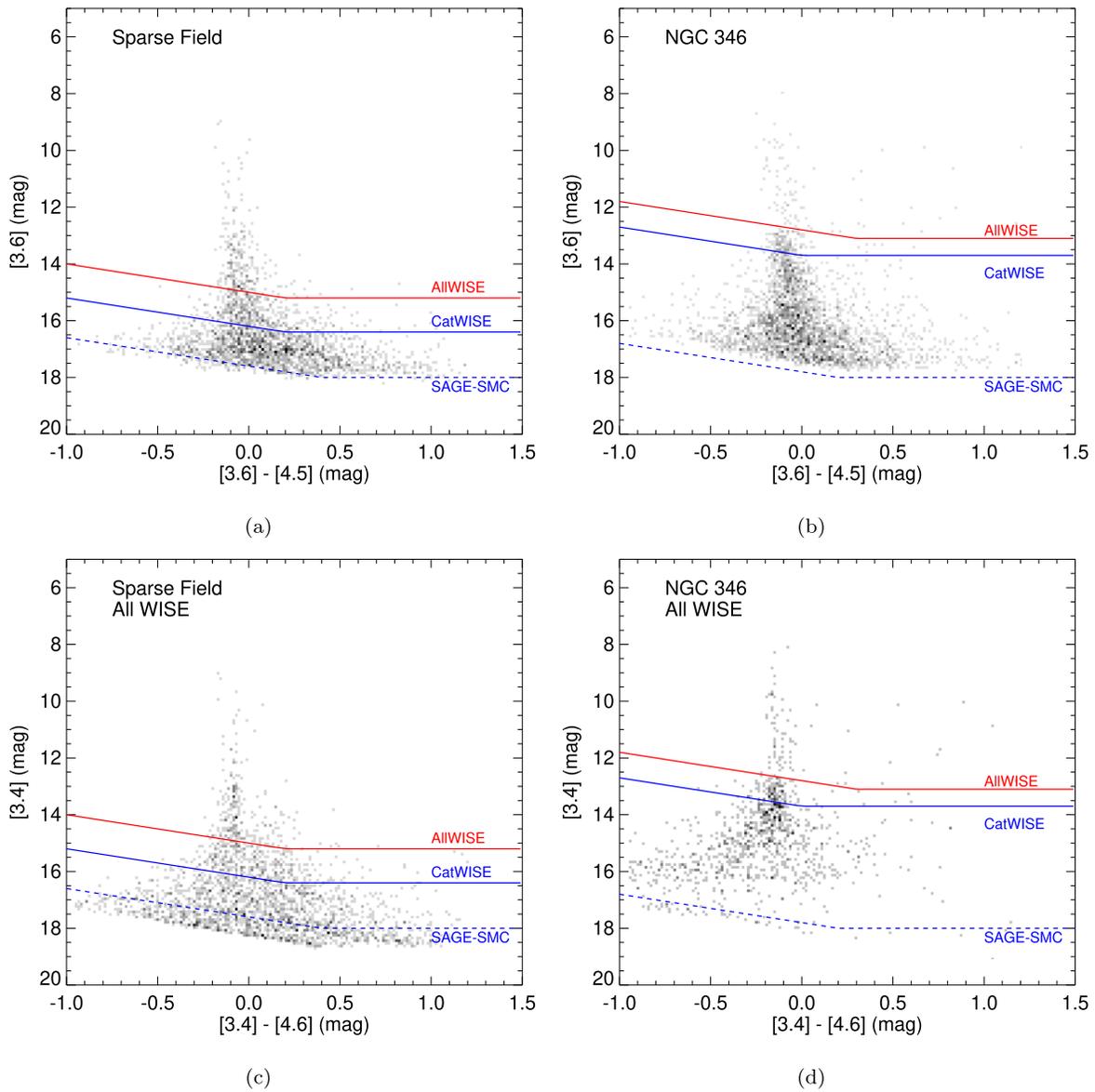

**Figure 8.** Color–magnitude diagrams of the two test fields for the epoch 1 sources in the SMC-Last archive (panels (a) and (b)) and for AllWISE (panels (c) and (d)). The lines in each plot represent the recovered magnitudes (for 95% matches) vs. color for the comparison catalogs as derived in Table 5: AllWISE (red), CatWISE (solid blue), and SAGE-SMC (dashed blue).

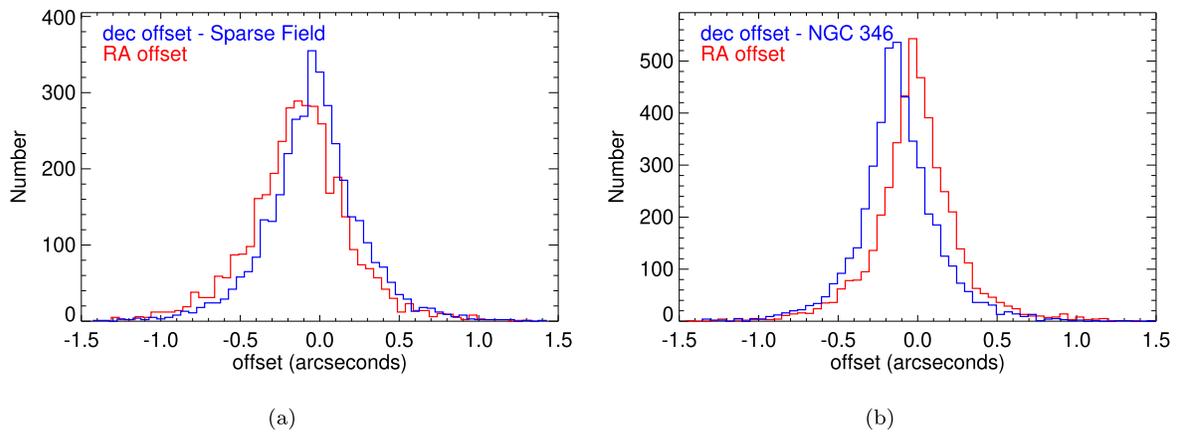

**Figure 9.** Positional differences in 0″.05 bins between the SMC-Last archive and SAGE-SMC archive for the two test fields. The differences are separated by R.A. (red) and decl. (blue).

Table 5
Comparison of Magnitude Limits in the Test Fields

| | NGC 346 Field | | | | | Sparse Field | | | | |
|---|---|---|---|---|---|---|---|---|---|---|
| | No. of Sources | 3.6 μm | | 4.5 μm | | No. of Sources | 3.6 μm | | 4.5 μm | |
| | | 95% | 90% | 95% | 90% | | 95% | 90% | 95% | 90% |
| SMC-Last Catalog | | | | | | | | | | |
| Epoch 1 | 1998 | 13.8 | 14.9 | 13.5 | 14.7 | 1473 | 15.8 | 16.2 | 15.6 | 16.0 |
| Epoch 2 | 2253 | 14.6 | 15.2 | 14.1 | 14.9 | 1299 | 15.8 | 16.1 | 15.5 | 16.0 |
| Combined | 3057 | 14.2 | 15.2 | 14.0 | 14.9 | 2005 | 16.0 | 16.6 | 15.8 | 16.3 |
| SMC-Last Archive | | | | | | | | | | |
| Epoch 1 | 4897 | 15.5 | 16.4 | 15.2 | 16.2 | 4291 | 17.2 | 17.5 | 16.9 | 17.2 |
| Epoch 2 | 4947 | 15.5 | 16.3 | 15.4 | 16.2 | 4067 | 17.1 | 17.4 | 16.8 | 17.0 |
| Combined | 5830 | 15.7 | 16.4 | 15.4 | 16.3 | 5697 | 17.6 | 18.0 | 17.1 | 17.6 |
| WISE Catalogs[a] | | | | | | | | | | |
| AllWISE | 1108 | 13.1 | 13.7 | 12.8 | 13.7 | 2421 | 15.2 | 16.0 | 15.0 | 15.9 |
| CatWISE2020 | 3020 | 13.7 | 14.7 | 13.7 | 14.6 | 5900 | 16.4 | 17.9 | 16.2 | 17.4 |
| SAGE-SMC | | | | | | | | | | |
| Archive | 7903 | 18.0 | 18.1 | 17.8 | 17.9 | 4659 | 18.0 | 18.1 | 17.6 | 17.7 |

**Note.**
[a] WISE magnitudes are reported for 3.4 and 4.6 μm.

99% of these sources were within $1''\!.5$ of each other. Figure 9 displays the positional differences for the two fields. Overall, the mean positional differences of the two fields were consistent, $0''\!.15 \pm 0''\!.08$ and $0''\!.17 \pm 0''\!.15$ for the NGC 346 and sparse fields, respectively. The differences in R.A. and decl. were $\Delta\alpha = -0''\!.01 \pm 0''\!.17$ and $\Delta\delta = -0''\!.13 \pm 0''\!.17$ in the crowded field and $\Delta\alpha = 0''\!.12 \pm 0''\!.23$ and $\Delta\delta = -0''\!.04 \pm 0''\!.22$ in the sparse field. These differences are consistent with the uncertainties estimated in Section 3.3, and they are also consistent with the astrometric uncertainty for SAGE-SMC of $0''\!.3$ (SAGE–SMC Team & Gordon 2011).

## 6. Summary

We have created a point-source catalog from the SMC-Last mosaics with roughly 1 million highly reliable sources with both 3.6 and 4.5 μm detections. A less reliable but more complete "archive" contains ∼2.4 million sources. Separate versions are provided for the two epochs in which SMC-Last surveyed the SMC and for a third, combined epoch. Combining the SMC-Last source lists with those from SAGE-SMC will enable a range of time-domain investigations with temporal baselines of nearly 10 yr throughout the surveyed region and over 12 yr in the core of the SMC. Adding the WISE catalogs from AllWISE and CatWISE will extend that to over 15 yr (and growing). Potential studies include searching for transient behavior in young stellar objects and cataclysmic variables and exploring the long-term variability in evolved stars. Foreground brown dwarfs can also be detected due to their proper motions and the long temporal baseline available. The data will also provide a mid-infrared complement to catalogs at other wavelengths, such as Gaia in the optical, and for potential missions in the near- and far-infrared (e.g., the Roman Telescope and possibly NASA's next Flagship mission).

The SMC-Last point sources are being delivered to the Infrared Science Archive (IRSA), where they will be available to the community along with the mosaic images.

## Acknowledgments

This work is based on observations made with the Spitzer Space Telescope, which was operated by the Jet Propulsion Laboratory, California Institute of Technology, under NASA contract 1407. We thank K. Gordon for helpful discussions in the observation planning and source extraction, and we thank Onemo Kang for assisting in initial quality checks for the source extractions. Financial support for this work was provided by NASA through NASA ADAP grant 80NSSC19K0585. M.S. acknowledges support from NASA ADAP grant No. 80NSSC22K0168. F.K. was supported by the Spanish program Unidad de Excelencia María de Maeztu CEX2020-001058-M, financed by MCIN/AEI/10.13039/501100011033. S.S. acknowledges support from UNAM-PAPIIT Program IA 104822. This research made use of data products from the Wide-field Infrared Survey Explorer, which is a joint project of the University of California, Los Angeles, and the Jet Propulsion Laboratory/California Institute of Technology, funded by the National Aeronautics and Space Administration. Based in part on data products created from observations collected at the European Organisation for Astronomical Research in the Southern Hemisphere under ESO program 179.B-2003. This research used the NASA/IPAC Infrared Science Archive, which is funded by the National Aeronautics and Space Administration and operated by the California Institute of Technology, NASA's Astrophysics Data System, and CDS's Vizier and Simbad services.

*Facilities:* Spitzer (IRAC), WISE, IRSA.

## Appendix A
## Saturated Sources

Table 6 lists the 19 very bright sources that were moved from the catalog to the archive. Most sources are within $0''\!.5$ of the Simbad coordinates of the object listed in the "Other Name" column. Four sources have larger offsets: HD 6623 ($1''\!.9$), κ

Table 6
Very Bright Sources

| SMC-Last Name | R.A. (J2000) | Decl. (J2000) | Other Name |
|---|---|---|---|
| J005752.24−702730.3 | 14.467667 | −70.458405 | CY Tuc |
| J010431.29−703225.2 | 16.130362 | −70.540344 | CZ Tuc |
| J012746.95−705126.3 | 21.945627 | −70.857292 | HD 9162 |
| J005412.88−713704.4 | 13.553684 | −71.617897 | Flo 286 |
| J010508.31−714401.5 | 16.284643 | −71.733742 | HD 6623[a] |
| J011100.45−713629.4 | 17.751877 | −71.608170 | PMMR 191 |
| J012738.94−714214.1 | 21.912251 | −71.703903 | HD 9163 |
| J010053.75−724151.3 | 15.223945 | −72.697578 | HD 6172 |
| J010905.44−723833.2 | 17.272650 | −72.642555 | HD 7100 |
| J005553.58−731827.3 | 13.973235 | −73.307594 | CM Tuc |
| J020158.15−733001.6 | 30.492310 | −73.500443 | HD 12841 |
| J022251.93−733844.7 | 35.716362 | −73.645752 | κ Hyi[a] |
| J003545.73−735240.9 | 8.940524 | −73.878036 | HD 3407 |
| J005500.29−741806.6 | 13.751199 | −74.301826 | HD 5499[a] |
| J020025.93−743701.5 | 30.108030 | −74.617081 | HD 12714 |
| J021117.66−743004.8 | 32.823570 | −74.501343 | HD 13950 |
| J004835.93−745524.8 | 12.149715 | −74.923561 | λ Hyi[a] |
| J012744.88−750236.2 | 21.937010 | −75.043381 | HD 9248 |
| J015653.88−752156.1 | 29.224518 | −75.365578 | HD 12362 |

**Note.**
[a] Stars with Simbad matches >1″.

Hyi (1″.6), HD 5499 (2″.8), and λ Hyi (2″.1). All four have high proper motions, between ∼70 and 170 mas yr$^{-1}$ (Gaia Collaboration et al. 2023), which likely caused the positional offsets.

## Appendix B
## Pointing Refinements

Nearly all of the astronomical observing requests (AORs) in the SMC-Last survey contain the "superboresight" pointing solution as the nominal astrometry information in the headers of the basic calibrated data (BCD) products (i.e., the CRVAL1 and CRVAL2 values), which the IRAC instrument handbook indicates should have an rms accuracy of about 0″.16. However, we found that the data contain significantly worse errors, primarily in the R.A., over the entire data set. Comparing the SMC-Last source positions extracted from the data frames with matching source coordinates in the 2MASS 6× catalog (Cutri et al. 2012), we found a quasiperiodic error in the R.A., with the errors characteristic of individual frames, with an approximate period of 80 s and a peak-to-peak amplitude of about 0″.7, although the period and amplitude are not constant. The pattern of errors is nearly identical for both bands. Figure 10 shows the R.A. errors for the first 150 10.4 s frames for AORKEY 64019968 for both bands. The absolute errors have been calculated by mapping 2MASS 6× sources into the astrometry for each data frame, matching to the sources in the frame array, and taking the median of the differences in R.A. over all the matched sources for that frame. (Figure 10 also shows the decl. errors, but they generally fall within the advertised superboresight errors.)

In addition, the "superboresight" solution was not included in the corrected BCD products for the first six AORs of the second epoch (covering the core of the SMC). For these AORs, the decl. as well as the R.A. show similar error patterns, plus global errors on the order of an arcsecond. In these cases, the error patterns in R.A. and decl. are correlated, although the global errors are different between the two coordinates.

For most of our AORs, a "refined" pointing solution has been included in the headers of the BCD products supplied by IPAC, in which the center R.A. and decl. are assigned values from matches of the sources on each data frame to sources in the 2MASS catalog. These refined coordinates show a considerable improvement in accuracy over the "superboresight" coordinates. However, not all of our AORs include these refined coordinates, and so we have developed a pointing refinement procedure and applied it to all of the SMC-Last data.

We use the 2MASS catalog as our "truth" reference (Cutri et al. 2012). The 6× catalog covers about three-fourths of our survey region, including the core of the SMC, and omits only the tail eastward of about 01$^h$52$^m$. For this portion, we use the regular 2MASS catalog sources (Skrutskie et al. 2006). We use for the refinement only sources with K brighter than 18th magnitude. This limit corresponds to sources of ∼0.025 mJy in the 3.6 μm band, which typically have an S/N of about 4 in the frame images (∼3 for the 4.5 μm band), so we expect that the vast majority of included 2MASS sources will correspond to a detectable source in our data.

For each data frame, the 2MASS sources are mapped into the array, using the nominal astrometry in the frame header. For each source, the peak pixel in a 5 × 5 pixel box centered on the mapped pixel is found, and a simple centroiding procedure is then applied: a 5 × 5 pixel box about the peak pixel is considered, and if there are any masked or missing-data pixels in the box, the source is ignored. The perimeter median of the 5 × 5 pixel box is subtracted as the background, and any resulting negative pixels are set to 0. The centroid of the 3 × 3 pixel box centered on the peak is calculated. The centroid is then converted to R.A. and decl. with the nominal frame astrometry, and the coordinate error for that source is determined from the "truth" 2MASS coordinates. The median R.A. and decl. errors over the 2MASS sources on the frame are then applied as a correction to the CRVAL1 and CRVAL2 values. We have not attempted to determine a correction for any field rotation or scaling, and so the Coordinate Description matrix values in the headers are not adjusted.

That the R.A. errors are nearly identical for the contemporaneous data from the frames for both bands suggests that the correction is essentially a boresight correction. It should be advantageous to combine the errors in the two bands for better statistics, particularly for sparse regions with the fewest 2MASS sources per frame. However, that would introduce a complication. Figure 11 shows a small but clear offset in the R. A. errors between the two bands; the median over all frames in this AOR is about 0″.066. These small correction offsets, with varying values, are seen in all the data, including the decl., so this effect seems independent of the patterned error seen in the RA. Examining these errors for AORs observed with various roll angles reveals that these offsets are consistent with a displacement fixed in focal plane coordinates (i.e., x and y on the array), with a magnitude of about 0″.07 in a direction about 37° counterclockwise from the +y-axis. We speculate that this offset is due to a small inaccuracy in the positioning of the arrays on the focal plane.

Regardless of the cause, we can use these results to determine corrections from the combined errors in both bands. For cases with at least 50 2MASS sources on each of the arrays in both bands, the separate coordinate corrections are applied as-is. This threshold is somewhat arbitrary, but with an rms of about $0''\!.5$ for the centroid precision in the 2MASS matches, the expected residual error for 50 sources is about $0''\!.07$, well under the typical positional scatter in the source extractions from the mosaics (Figure 4), so combining the error statistics for both bands for this number of sources (or greater) would have little benefit.

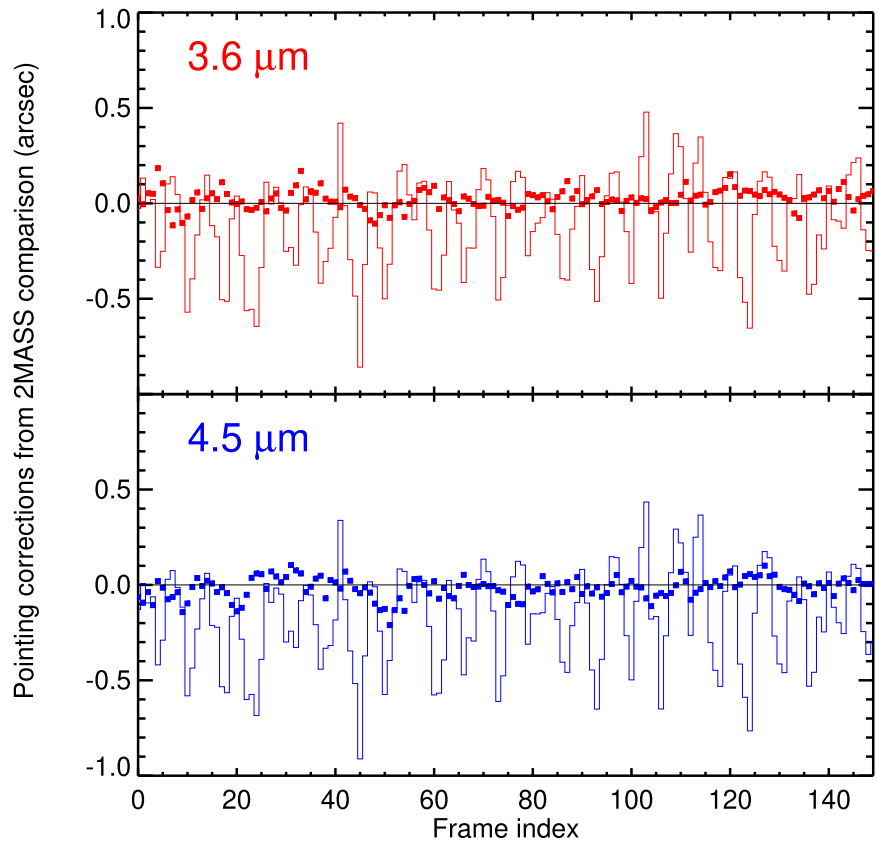

**Figure 10.** Frame-by-frame pointing corrections from 2MASS source comparisons for the first 150 10.4 s frames of AORKEY 64019968, showing the contemporaneous 3.6 and 4.5 $\mu$m corrections. R.A. corrections are plotted as lines and decl. corrections as dots.

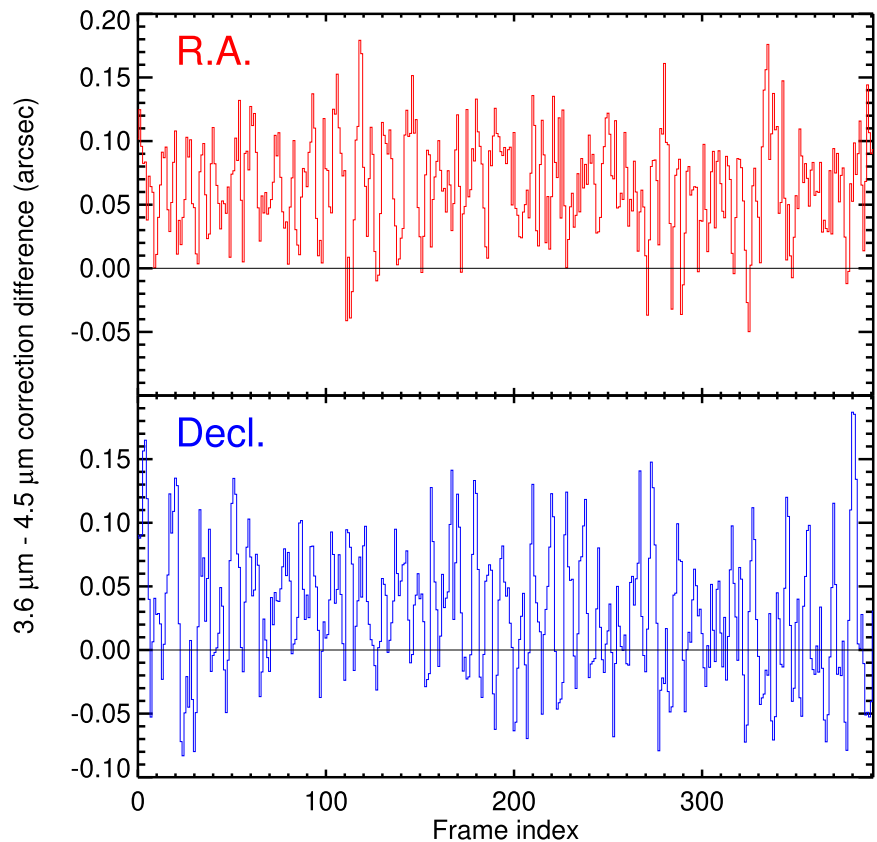

**Figure 11.** The difference between the 3.6 and 4.5 $\mu$m pointing corrections for the 392 10.4 s frames in AORKEY 64019968.

For fewer than 50 sources in either band, the medians of the R.A. and decl. errors are taken over both arrays, and the fixed focal plane offset is converted to offsets in R.A. and decl. and then apportioned to the separate arrays as an adjustment to the overall correction depending on the fraction of the total sources on each array. This procedure requires that the separate R.A. and decl. error distributions for the two arrays sufficiently overlap so that the median is close to the average, and this is in fact the case (the scatter of the individual errors is around half an arcsecond).

Finally, with the pointing refinement applied as above, it was found that the subsequent source extractor coordinates measured from mosaics created from the corrected frames differed systematically from the 2MASS coordinates by a small amount, about $0''\!.055$, also fixed in focal plane coordinates, which is likely due small differences in the centroiding procedures used and possibly the slightly asymmetric PRFs. As the extractor coordinates are the ones ultimately reported, this offset is also added to the calculated pointing corrections.

### B.1. Scaling and Rotation

The corrections applied are translations in the focal plane, because the data show that the pointing errors are largely time-varying errors in the boresight, which for reasonable magnitudes of error will only cause translations in the $5'$-size IRAC arrays. For that reason, it is highly unlikely that there will be any true frame-by-frame variations in scaling or field rotation. Thus, it is reasonable to investigate the presence of fixed errors in scaling and rotation in the IRAC frame astrometry.

For this purpose, we have developed a $\chi^2$ minimization procedure to optimize the mapping of the measured $x$, $y$ centroids of the sources on a given frame to the $x$, $y$ coordinates of the corresponding 2MASS sources, using the frame's nominal astrometry to map the 2MASS sources into focal plane space. Five parameters are optimized: $x$ and $y$ scaling, a rotation, and $x$ and $y$ translation.

Optimized parameters are determined for each IRAC frame of an AOR (using AORs in fairly crowded regions for better statistics), then examined across the AOR for persistence. While there is considerable frame-to-frame scatter, particularly in the rotation, we find no significant systematic errors in the scaling parameters or the 4.5 $\mu$m rotation parameter in any of the AORs examined. However, the optimized 3.6 $\mu$m rotation shows a median error of about $1'$, and this error is present at a nearly identical value over all of the AORs. A $1'$ field rotation error gives a peak true angular error of about $0''\!.07$ in the $5'$ IRAC arrays and an effective rms error of about $0''\!.03$ for an ensemble of sources distributed across the array.

## ORCID iDs

T. A. Kuchar ● https://orcid.org/0000-0003-1955-8509
G. C. Sloan ● https://orcid.org/0000-0003-4520-1044
D. R. Mizuno ● https://orcid.org/0000-0003-0947-2824
Kathleen E. Kraemer ● https://orcid.org/0000-0002-2626-7155
M. L. Boyer ● https://orcid.org/0000-0003-4850-9589
Martin A. T. Groenewegen ● https://orcid.org/0000-0003-2723-6075
O. C. Jones ● https://orcid.org/0000-0003-4870-5547
F. Kemper ● https://orcid.org/0000-0003-2743-8240
Iain McDonald ● https://orcid.org/0000-0003-0356-0655
Joana M. Oliveira ● https://orcid.org/0000-0002-0861-7094
Marta Sewiło ● https://orcid.org/0000-0003-2248-6032
Sundar Srinivasan ● https://orcid.org/0000-0002-2996-305X
Jacco Th. van Loon ● https://orcid.org/0000-0002-1272-3017
Albert Zijlstra ● https://orcid.org/0000-0002-3171-5469